\documentclass[12pt]{article}
\pdfoutput=1

\usepackage{amsmath,amssymb}
\usepackage{epsfig}

\usepackage{graphicx, rotating}
\usepackage{hyperref}
\usepackage{slashed}
\usepackage{ifpdf}
\usepackage{amsmath}
\usepackage{slashed}
\usepackage{cite}
\usepackage{multicol}
\usepackage{color}

\ifx\pdfoutput\undefined
\usepackage[dvips,bookmarks=false]{hyperref}	
\else
\usepackage{hyperref}	
\fi
\hypersetup{colorlinks,bookmarksopen,bookmarksnumbered,citecolor=verdes,
linkcolor=blus,pdfstartview=FitH,urlcolor=rossos}
\def\hhref#1{\href{http://arxiv.org/abs/#1}{#1}} 

\newcommand{\beq}{\begin{equation}}
\newcommand{\eeq}{\end{equation}}

\newcommand{\be}{\begin{equation}}
\newcommand{\ee}{\end{equation}}

\newcount\Mac  \Mac=1  
\newcommand{\ifMac}[2]{\ifnum\Mac=1 #1 \else #2 \fi}
\def\putps(#1,#2)(#3,#4)#5#6{\ifnum\Mac=1 \put(#1,#2){\special{picture #5}}
\else  \put(#3,#4){\includegraphics{#6}} \fi}

\newcommand{\One}{\hbox{1\kern-.24em I}}

\newcommand{\GeV}{\,{\rm GeV}}

\newcommand{\lascia}[1]{}
\makeatletter
\def\art{\@ifnextchar[{\eart}{\oart}}
\def\eart[#1]#2#3#4#5#6{{\rm #2}, {#3 #4} {\rm (#6) #5} [arXiv:{\hhref{#1}}]}
\def\hepart[#1]#2{{\rm #2, arXiv:\hhref{#1}}}
\newcommand{\oart}[5]{{\rm #1}, {#2 #3} {\rm (#5) #4}}

\newcounter{alphaequation}[equation]
\def\thealphaequation{\theequation\hbox to
0.6em{\hfil\alph{alphaequation}\hfil}}
\def\eqnsystem#1{
\def\@eqnnum{{\rm (\thealphaequation)}}
\def\@@eqncr{\let\@tempa\relax \ifcase\@eqcnt \def\@tempa{& & &} \or
  \def\@tempa{& &}\or \def\@tempa{&}\fi\@tempa
  \if@eqnsw\@eqnnum\refstepcounter{alphaequation}\fi
\global\@eqnswtrue\global\@eqcnt=0\cr}
\refstepcounter{equation} \let\@currentlabel\theequation \def\@tempb{#1}
\ifx\@tempb\empty\else\label{#1}\fi
\refstepcounter{alphaequation}
\let\@currentlabel\thealphaequation
\global\@eqnswtrue\global\@eqcnt=0 \tabskip\@centering\let\\=\@eqncr
$$\halign to \displaywidth\bgroup \@eqnsel\hskip\@centering
$\displaystyle\tabskip\z@{##}$&\global\@eqcnt\@ne
\hskip2\arraycolsep\hfil${##}$\hfil& \global\@eqcnt\tw@\hskip2\arraycolsep
$\displaystyle\tabskip\z@{##}$\hfil
\tabskip\@centering&\llap{##}\tabskip\z@\cr}

\def\endeqnsystem{\@@eqncr\egroup$$\global\@ignoretrue} \makeatother

\def\circa#1{\,\raise.3ex\hbox{$#1$\kern-.75em\lower1ex\hbox{$\sim$}}\,}

\definecolor{rosso}{cmyk}{0,1,1,0.4}
\definecolor{rossos}{cmyk}{0,1,1,0.55}
\definecolor{rossos2}{rgb}{0.8,0.2,0.3}
\definecolor{rossoc}{cmyk}{0,1,1,0.2}
\definecolor{blu}{cmyk}{1,1,0,0.3}
\definecolor{blus}{cmyk}{1,1,0,0.6}
\definecolor{bluc}{cmyk}{1,1,0,0.1}
\definecolor{verde}{cmyk}{0.92,0,0.59,0.25}
\definecolor{verdec}{cmyk}{0.92,0,0.59,0.15}
\definecolor{verdes}{cmyk}{0.92,0,0.59,0.4}
\definecolor{grigio}{cmyk}{0,0,0,0.07}
\definecolor{rosa}{cmyk}{0,0.1,0.1,0.02}
\definecolor{rosino}{cmyk}{0,0.05,0.05,0.02}
\definecolor{rosas}{cmyk}{0,0.3,0.25,0.05}
\definecolor{celeste}{cmyk}{0.1,0,0,0.02}
\definecolor{giallino}{cmyk}{0,0,0.4,0.02}

\font\tenrsfs=rsfs10 at 12pt

\font\sevenrsfs=rsfs7
\font\fiversfs=rsfs5
\newfam\rsfsfam
\textfont\rsfsfam=\tenrsfs
\scriptfont\rsfsfam=\sevenrsfs
\scriptscriptfont\rsfsfam=\fiversfs
\def\mathscr#1{{\fam\rsfsfam\relax#1}}

\def\beq{\begin{equation}}
\def\eeq{\end{equation}}
\def\bea{\begin{eqnarray}}
\def\eea{\end{eqnarray}}

\def\m@th{\mathsurround=0pt }
\def\leftrightarrowfill{$\m@th \mathord\leftarrow \mkern-6mu
        \cleaders\hbox{$\mkern-2mu \mathord- \mkern-2mu$}\hfill
        \mkern-6mu \mathord\rightarrow$}

\def\overleftrightarrow#1{\vbox{\ialign{##\crcr
        \leftrightarrowfill\crcr\noalign{\kern-1pt\nointerlineskip}
        $\hfil\displaystyle{#1}\hfil$\crcr}}}

\def\shat{\ifmmode \hat{s}\else $\hat{s}$\fi}
\def\gp2{{g'}^2}
\def\g2{g^2}
\def\g32{g_s^2}

\newcommand{\newc}{\newcommand}

\newc{\gsim}{\lower.7ex\hbox{$\;\stackrel{\textstyle>}{\sim}\;$}}
\newc{\lsim}{\lower.7ex\hbox{$\;\stackrel{\textstyle<}{\sim}\;$}}
\newc{\ie}{{\it i.e.}}
\newc{\etal}{{\it et al.}}
\newc{\mev}{\hbox{\rm\,MeV}}
\newc{\gev}{\hbox{\rm\,GeV}}
\newc{\tev}{\hbox{\rm\,TeV}}
\newc{\xpb}{\hbox{\rm\, pb}}
\newc{\xfb}{\hbox{\rm\, fb}}

\newc{\G}{{\cal G}}
\newc{\h}{{\cal H}}
\newc{\D}{{\cal D}}
\newc{\E}{{\cal E}}

\newc{\mtop}{m_t}
\newc{\mbot}{m_b}
\newc{\mz}{M_Z}
\newc{\mw}{M_W}
\newc{\alphasmz}{\alpha_s(M_Z)}
\newc{\swsq}{\sin^2\theta_W}
\newc{\cwsq}{\cos^2\theta_W}
\newc{\tw}{\tan\theta_W}
\newc{\cw}{\cos\theta_W}
\newc{\sw}{\sin\theta_W}
\newc{\BR}{\hbox{\rm BR}}
\newc{\zbb}{Z\to b\bar}
\newc{\Gb}{\Gamma (Z\to b\bar b)}
\newc{\Gh}{\Gamma (Z\to \hbox{\rm hadrons})}
\newc{\sgn}{\mbox{sgn}}

\newcounter{mysubequation}[equation]

\def\beq{\begin{equation}}
\def\eeq{\end{equation}}
\def\bea{\begin{eqnarray}}
\def\eea{\end{eqnarray}}
\def\slashchar#1{\setbox0=\hbox{$#1$}           
   \dimen0=\wd0                                 
   \setbox1=\hbox{/} \dimen1=\wd1               
   \ifdim\dimen0>\dimen1                        
      \rlap{\hbox to \dimen0{\hfil/\hfil}}      
      #1                                        
   \else                                        
      \rlap{\hbox to \dimen1{\hfil$#1$\hfil}}   
      /                                         
   \fi}                                         %
\catcode`@=11
\long\def\@caption#1[#2]#3{\par\addcontentsline{\csname
  ext@#1\endcsname}{#1}{\protect\numberline{\csname
  the#1\endcsname}{\ignorespaces #2}}\begingroup
    \small
    \@parboxrestore
    \@makecaption{\csname fnum@#1\endcsname}{\ignorespaces #3}\par
  \endgroup}
\catcode`@=12

\begin{document}

\color{black}

\begin{flushright} SISSA 24/2012/EP\end{flushright}

\vspace{1cm}
\begin{center}
{\Huge\bf\color{black} Cosmological Perturbations\\[0.2cm] from the Standard Model Higgs}\\
\bigskip\color{black}\vspace{1cm}{
{\large{Andrea De Simone} $^{\rm {a,b}}$ 
and {Antonio Riotto} $^{\rm c}$}
\vspace{0.8cm}
} \\[7mm]
{\em $^{\rm a}$ {SISSA, Via Bonomea 265, I-34136 Trieste, Italy}}\\
{\em $^{\rm b}$ {INFN, Sezione di Trieste, I-34136 Trieste, Italy}}\\
{\em $^{\rm c}$ {Department of Theoretical Physics and Center for Astroparticle Physics (CAP),\\ 24 quai E. Ansermet, CH-1211 Geneva 4, Switzerland}}\\
\end{center}
\bigskip
\centerline{\large\bf Abstract}
\begin{quote}\small
We propose that the Standard Model (SM) Higgs  is responsible for generating the cosmological perturbations of the universe by acting as an isocurvature mode during a de Sitter inflationary stage. In view of the recent ATLAS and CMS results for the   Higgs mass, this can happen if the Hubble rate during inflation is in the range $(10^{10}- 10^{14})$ GeV (depending on the SM parameters).
Implications for the detection of primordial tensor perturbations through the $B$-mode of CMB polarization via the PLANCK satellite are discussed. For example, 
if the Higgs mass value is confirmed
to be $m_h=125.5$ GeV and $m_t, \alpha_s$ are at their central values, our mechanism predicts tensor perturbations too small
to be detected in the near future. 
On the other hand, if tensor perturbations will be detected by PLANCK through the $B$-mode of CMB, then 
there is a definite relation between the Higgs and top masses,  
making the mechanism predictive and falsifiable.

\end{quote}

\normalsize



\def\thefootnote{\arabic{footnote}}
\setcounter{footnote}{0}
\pagestyle{empty}

\newpage
\pagestyle{plain}
\setcounter{page}{1}

\section{Introduction}
\noindent
Experimental data recently reported by ATLAS \cite{ATLAS1,ATLAS2} and CMS \cite{CMS1,CMS2} 
are consistent with the discovery of  the  Standard Model (SM) Higgs boson, with  mass 
around $125-126$  GeV.
 
Such a light Higgs is in good agreement with the indirect indications 
derived from electroweak precisions constraints~\cite{EW} under the hypothesis of negligible 
contributions of physics beyond the SM. Moreover, no clear signal of non-SM physics
has emerged yet from collider searches. 

Motivated by this experimental situation, in the present paper we try to answer a simple question: 
can the SM Higgs be responsible for the cosmological perturbations we observe in the universe? As we will see, the answer is: yes.

One of the basic ideas of modern cosmology is that there was an epoch early
in the history of the universe when potential, or vacuum, energy 
associated to a scalar field, the inflaton, 
dominated other forms of energy density such as matter or radiation. 
During such a
vacuum-dominated era the scale factor grew (nearly) exponentially
in time. During this phase, dubbed inflation 
\cite{guth81,Starobinsky, lrreview},
a small,  smooth spatial region of size less than the Hubble radius
could grow so large as to easily encompass the comoving volume of the 
entire presently observable universe. If the universe underwent
such a period of rapid expansion, one can understand why the observed
universe is so homogeneous and isotropic to high accuracy.

Inflation has also become the dominant 
paradigm for understanding the 
initial conditions for the Large Scale Structure (LSS) formation and for Cosmic
Microwave Background (CMB) anisotropy. In the
inflationary picture, primordial density and gravity-wave fluctuations are
created from quantum fluctuations ``redshifted'' out of the horizon during an
early period of superluminal expansion of the universe, where they
are ``frozen'' \cite{starob79, muk81,hawking82,starobinsky82,guth82,bardeen83}. 
Perturbations at the surface of last scattering are observable as temperature 
anisotropy in the CMB. 
The last and most impressive confirmation of the inflationary paradigm has 
been recently provided by the data 
of the Wilkinson Microwave Anistropy Probe (WMAP) mission which has 
marked the beginning of the precision era of the CMB measurements in space
\cite{wmap7}.

Despite the simplicity of the inflationary paradigm, the mechanism
by which  cosmological adiabatic perturbations are generated  is not
yet fully established. In the standard picture, the observed density 
perturbations are due to fluctuations of the inflaton field itself. 
When inflation ends, the inflaton oscillates about the minimum of its
potential and decays, thereby reheating the universe. As a result of the 
fluctuations
each region of the universe goes through the same history but at slightly
different times. The 
final temperature anisotropies are caused by the fact that
inflation lasts different amounts of time in different regions of the universe
leading to adiabatic perturbations. 

Can the  SM Higgs and its potential
be responsible for inflation
 and, at the same time,  the generation of anisotropies? 

 The answer is: most probably,  not both
\cite{ry}.  The
basic problem is that the requirement of having enough e-folds of inflation requires the SM potential to be flat enough, but  this 
conflicts with the requirement that quantum fluctuation
of the Higgs inflaton should also generate the observed power spectrum of anisotropies. Indeed the height  of the SM potential in its 
flat region is predicted and cannot be arbitrarily adjusted
to be as low as needed. This problem can be, in principle, solved by non-minimal coupling of the SM Higgs scalar field $h$  to the Ricci scalar
$R$ of the form $\xi h^2 R$  \cite{hi,hi1,Barvinsky}. The effect of this interaction is to flatten the Higgs
potential (or any other potential) above the  scale $M_{\rm Pl}/\sqrt{\xi}$  providing a platform for slow-roll
inflation. A correct normalization of the spectrum of primordial 
fluctuations fixes the value of the coupling constant $\xi$ to be larger than about  $10^4$. 

This minimal inflationary scenario faces though a couple of issues.
First, perturbative unitarity is violated at some scale lower than $M_{\rm Pl}$ \cite{hi2}
(see however Ref.~\cite{sibi}), 
possibly implying the presence of new degrees of freedom which may change the inflationary
dynamics. 
Secondly,  the scenario requires stability of the potential up to
the inflationary scale  $M_{\rm Pl}/\sqrt{\xi}$. 
 With  the LHC indication of a Higgs mass in the range  $m_h = (125-126)$  GeV,
this simplest version of Higgs inflation is disfavored as the Higgs potential develops an instability at much smaller scales, unless the top mass is below $\sim 171$ GeV \cite{hi3,hi4,hi5,Bezrukov:2012sa}.

These considerations lead us to believe that the SM Higgs field may not be responsible for both driving inflation and generating
the cosmological perturbations at the same time. Let us therefore be more modest and drop off the requirement that the SM Higgs potential
was responsible for inflation.  Our goal is to show that 
 the SM Higgs can nevertheless  play a role in giving rise to the LSS and CMB anisotropies. Indeed, 
 the standard scenario for the
 generation of the perturbations, where it is the  same scalar field to drive inflation and to source the perturbations,  is not the only option.       
For instance, an alternative to the standard scenario is represented by the curvaton 
mechanism
\cite{Linde:1996gt, curvaton1,LW,curvaton3} where the final curvature perturbation $\zeta$
is produced from an initial isocurvature perturbation associated to the
quantum fluctuations of a light scalar field (other than the inflaton), 
the curvaton, whose energy density is negligible during inflation. The 
curvaton isocurvature perturbations are transformed into adiabatic
ones when the curvaton decays into radiation much after the end 
of inflation. Alternatives to the curvaton model are
those models characterized by the curvature perturbation being generated by an inhomogeneity in
the decay rate \cite{rate} or the mass \cite{mass} of the particles responsible for the reheating after inflation.
Other opportunities for generating the curvature perturbation occur at the end of inflation \cite{end} and
during preheating \cite{during}

All these alternative models to generate the cosmological perturbations have their strength in the fact that all scalar fields
during a period of de Sitter with a mass smaller than the Hubble rate $H$ during inflation are inevitably quantum-mechanically excited with a final superhorizon flat spectrum. Furthermore, they have 
in common that the comoving curvature perturbation is
 generated on superhorizon scales when the isocurvature perturbation, which is  associated to the fluctuations of these light scalar fields different from the inflaton,  is converted
into curvature perturbation after (or at the end) of inflation.

In the rest of the paper we will therefore assume that there was an inflationary period of accelerated expansion during the primordial
evolutionary stage of the universe.  This de Sitter period, induced by some unspecified vacuum energy,  is characterized by a Hubble rate $H$. We will also assume that the perturbations are
generated through the Higgs field. This will allow us to play with two independent parameters, the SM Higgs mass $m_h$ and the Hubble rate $H$.  Also, for simplicity, we assume that the inflaton sector does not alter the SM sector, {\it e.g.} the Higgs potential
(see also Ref.~\cite{cai} for an alternative idea where the Higgs sector of the SM is minimally coupled
to asymptotically safe gravity). 

Our  considerations will   have direct observational consequences once one realizes
that the Hubble rate 
parametrizes 
the amount of tensor perturbations during inflation. During
the inflationary epoch, tensor perturbations, as for any other massless 
scalar
field, are quantum-mechanically generated. They can give rise to $B$-modes
of polarization of the CMB radiation
through Thomson scatterings of the CMB photons off free electrons
at last scattering \cite{pol}. The amplitude of the $B$-modes
depends on the amplitude of the gravity waves
generated during inflation, which in turn depends on the
energy scale at which inflation occured. The tensor-to-scalar power ratio
 is given by $T/S \simeq (H/3.0\times 10^{14}\,
{\rm GeV})^2$. Current CMB anisotropy data impose the upper bound
$T/S\lsim 0.24$ \cite{wmapping}.
The possibility of detecting gravity waves from inflation via
$B$-modes is currently being considered by a number of ground, balloon and
space based experiments, included the PLANCK experiment. The decomposition of the CMB polarization into
$E$- and $B$-modes requires a full sky data coverage and, as such, is limited
by the foreground contaminations. The latter introduce a mixing of the
$E$ polarization into   $B$ with the corresponding cosmic variance limitation.
PLANCK's expected sensitivity is  about $T/S=0.05$ corresponding to a minimum testable value of    $H\simeq 
6.7\times 10^{13}$ GeV \cite{plancktensor}.
A detection of the tensor mode would therefore imply that the value of the Hubble rate during inflation is  
larger than
about  $10^{13}$ GeV. As we will see, this will have important implications for the the ideas discussed in this paper

The paper is organized as follows.
In Section \ref{sec:perturbation} we describe some  mechanisms by which the SM Higgs could generate the cosmological perturbation. Being ignorant about the exact mechanism, we try to be as generic as possible.
In Section \ref{sec:results} we  present our results and draw our conclusions.

\section{Curvature perturbation from the SM Higgs}
\label{sec:perturbation}

\noindent
Let us see in more detail how the curvature perturbation may be originated from the fluctuation of the Higgs field.
A convenient and rather model-independent way to characterize the curvature perturbation generated during or after a de Sitter stage is by the $\delta N$ formalism \cite{deltaN, starobinsky82, Starobinsky:1986fxa}. The  comoving curvature perturbation $\zeta$ 
on a uniform
energy density hypersurface at time $t_{\rm f}$ is, on sufficiently large scales, equal to the perturbation
in the time integral of the local expansion from an initial flat hypersurface ($t = t_{*}$)
to the final uniform energy density hypersurface. On sufficiently large scales, the local expansion
can be approximated quite well by the expansion of the unperturbed Friedmann
universe. Hence the curvature perturbation at time $t_{\rm f}$ can be expressed in terms of  the values of the relevant light scalar
fields (the inflaton field, the SM Higgs, etc.) $\sigma^I(t_{*},\vec{x})$ at $t_{*}$
\be
\zeta(t_{\rm f},\vec{x})=N_I\sigma^I+\frac{1}{2}N_{IJ}\sigma^I\sigma^J+\cdots \,\,\,\, \,\,\,\, \,\,\,\,\,\,\,\, \,\,\,\, \,\,\,\,\,\,\,\, (I=1,\cdots, M),
\label{deltan}
\ee
where $N_I$ and $N_{IJ}$ are the first and second derivative, respectively, of the number of e-folds 
\be
N(t_{\rm f},t_{*},\vec{x})=\int_{t_{*}}^{t_{\rm f}}\,{\rm d}t\, H(t,\vec{x})
\ee
with respect to the 
field $\sigma^I$. From the expansion (\ref{deltan}) one can read off the
 the two-point  
correlator of the comoving curvature perturbation  in momentum space
\begin{eqnarray}
P_\zeta(k_1)&=&N_IN_JP^{IJ}_{\vec{k}_{1}}, \nonumber\\
\langle\sigma_{\vec{k}_{1}}^I\sigma^J_{\vec{k}_{2}}\rangle&=&(2\pi)^3\delta({\vec{k}_{1}}+{\vec{k}_{2}})P^{IJ}_{\vec{k}_{1}}=
(2\pi)^3\delta({\vec{k}_{1}}+{\vec{k}_{2}})\delta^{IJ}\left(\frac{H}{2\pi}\right)^2_{k_1}.
\end{eqnarray}
The last passage, stating that   perturbations are not cross-correlated, holds if the cosmological perturbation is sourced by light scalar fields other than the inflaton as a consequence of the spatial
conformal symmetry  enjoyed by the de Sitter geometry \cite{kr}. The subscript $_{k_1}$ is there to remind that perturbations have to be evaluated at horizon-crossing, when $k_1=aH$, being $a$ the scale factor.
From now on we restrict ourselves to the case of $M=2$ and we identify one of the two fields with the adiabatic inflaton mode $\phi$ and the other with the SM Higgs. We also assume here that the contribution from the inflaton is subleading, $N_\phi\ll N_h$.

A specific example where the primordial density perturbations may be produced just
after the end of inflation is the modulated decay scenario when the decay rate of the inflaton is a function of the SM Higgs field \cite{rate}, that is  $\Gamma=\Gamma(h)$.
If we approximate the inflaton reheating by a sudden decay, we may find an analytic estimate of the density perturbation.
In the case of modulated reheating, the decay occurs on a spatial hypersurface with
variable local decay rate and hence local Hubble rate $H=\Gamma(h)$. Before the 
inflaton decay, the oscillating inflaton field has a pressureless equation of state and there is no density
perturbation. The perturbed expansion reads
\be
\delta N_{\rm d}=-\frac{1}{3}\ln\left(\frac{\rho_\phi}{\overline{\rho}_\phi}\right).
\ee
Immediately after the decay  we have radiation and hence the curvature perturbation reads
\be
\zeta=\delta N_{\rm d}+\frac{1}{4}\ln\left(\frac{\rho_\phi}{\overline{\rho}_\phi}\right).
\ee
Eliminating $\delta N_{\rm d}$ and using the local Friedmann equation $\rho\sim H^2$, to determine the local density in terms of the local decay rate $\Gamma=\Gamma(h)$, we have at the linear order
\be
\label{rate}
\zeta=-\frac{1}{6}\,\frac{\delta\Gamma}{\Gamma}=
-\frac{1}{6}\,\frac{{\rm d\ln }\Gamma}{{\rm d}\ln \overline{h}}\,\frac{\delta h}{\overline h}= -\beta_h\frac{\delta h}{\overline{h}}.
\ee
A maybe more intuitive way of understanding the expression (\ref{rate}) is to remember that, if the inflaton decay is perturbative, the
final reheating temperature $T_r$ scales like $(M_{\rm Pl} \Gamma)^{1/2}$ and therefore large scale spatial variations of the decay rate
will induce a temperature anisotropy, $\delta T_r/T_r\sim \delta\Gamma/\Gamma$.
The corresponding power spectrum of the curvature perturbation is  given by
\be
\label{aaaa}
P_{\zeta}= \beta_h^2\left(\frac{H}{2\pi\overline{h}}\right)^2.
\ee
Another possibility is that that the dominant component of
the curvature perturbation is generated at the transition between inflation and the post-inflationary phase \cite{end}.  
Let us suppose that slow-roll inflation suddenly
gives way to radiation domination through a waterfall transition in hybrid inflation and that the value of the inflaton $\phi_{\rm e}$ at which
this happens depends on the SM Higgs field, $\phi_{\rm e}=\phi_{\rm e}(h)$. If we assume again that the contribution to the
curvature perturbation from the inflaton field is subdominant, we get
\be
\zeta= N'_{\rm e}\delta \phi_{\rm e}=N'_{\rm e}\frac{{\rm d}\phi_{\rm e}}{{\rm d}\ln \overline{h}}\frac{\delta h}{\overline{h}}=
\frac{{\rm d}N_{\rm e}}{{\rm d}\ln \overline{h}}\frac{\delta h}{\overline{h}}
\ee
and the power spectrum is given by Eq. (\ref{aaaa}) with $\beta_h={\rm d}N_{\rm e}/{\rm d}\ln \overline{h}$. Given our ignorance about the parameter  $\beta_h$, in the following  we will treat it as a free parameter.  Specific forms of the function $\beta_h$ may give rise to other interesting and observable properties, such as 
non-Gaussianity in the perturbation \cite{reviewNG}. We will come back to these issues  in the future.

In all the considerations made so far we have assumed that the Higgs field is quantum-mechanically excited during the de Sitter stage and acts like an isocurvature mode. This imposes various conditions:
\begin{enumerate}

\item  The Higgs field does not contribute significantly  to the energy density during inflation, $V(h)\ll H^2 M_{\rm Pl}^2$.

\item  The SM Higgs field is light enough during inflation, that is the second derivative of the SM Higgs potential $V(h)$ is smaller than the square of the Hubble rate, $|{\rm d}^2 V(h)/{\rm d} h^2|\ll H^2$. This condition has to hold for a number of e-folds  large enough to create a sufficiently  homogenous and isotropic region 
encompassing the comoving volume of the 
entire presently observable universe (we take 60 as a fiducial number) and also
implies  that the  spectral index $n_\zeta$ is sufficiently close to unity
\be
n_{\zeta}-1=
\frac{{\rm d} {\rm ln} \,P_{\zeta}}{{\rm d} {\rm ln} \,k}=
\frac{{\rm d} {\rm ln} \,H_k^2}{{\rm d} {\rm ln} \, k}+\frac{2}{3 H^2}\frac{{\rm d}^2 V(h)}{{\rm d} h^2}
=-2\epsilon +\frac{2}{3 H^2}\frac{{\rm d}^2 V(h)}{{\rm d} h^2}.
\ee
We take $|{\rm d}^2 V(h)/{\rm d} h^2|/(3 H^2)< 10^{-2}$ as a sufficient starting condition. This also automatically implies that $|{\rm d} \ln h/{\rm d}\ln a |\ll 1$. 
\end{enumerate}

\section{Results and Conclusions}
\label{sec:results}

We use 2-loop renormalization group (RG) equations for all SM
couplings (gauge, Higgs quartic and top-yukawa couplings),
and the pole mass matching scheme for the Higgs and top masses,
as given in the Appendix of Ref.~\cite{hi3}.
The numerical solution of these equations allows to obtain the  RG-improved effective potential for the SM Higgs \cite{2loop}, as a function of input parameters, such as the Higgs and top masses.
For the top mass we considered $m_t=173.1\pm 0.7$ GeV (as in Ref.~\cite{hi5}
by combining Tevatron \cite{Aaltonen:2012ra} and LHC results), while for the QCD gauge coupling $\alpha_s(M_Z)=0.1184\pm 0.0007$ \cite{alphas}.
In all simulations we have set $\alpha_s$ to its central value; the size of
the effect of the variation of $\alpha_s$ within $1\sigma$ is comparable with the higher-order corrections (such as 3 loops) we are neglecting.

\begin{figure}[t]
\centering
\includegraphics[scale=1]{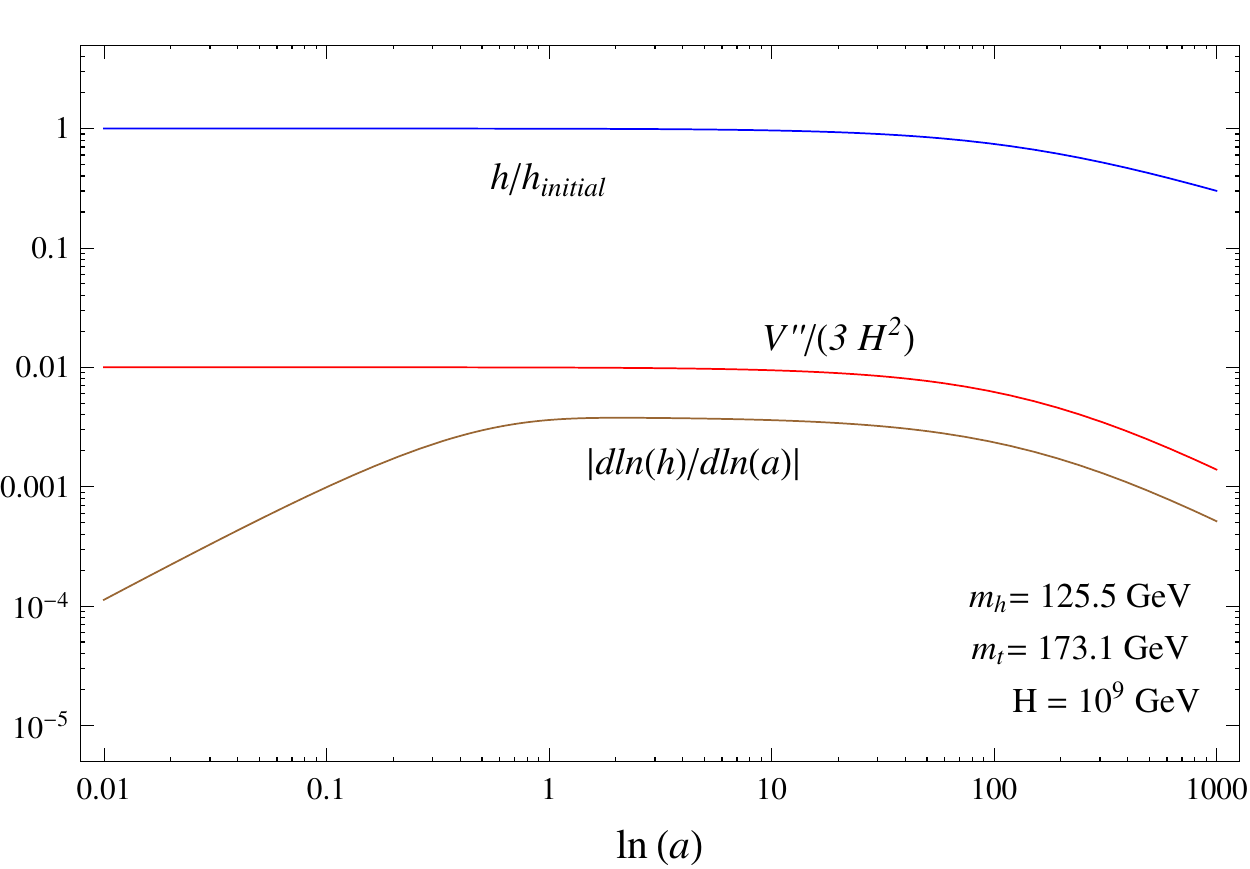}
\caption{The solution of the field equation (\ref{eq:fieldeq}) for $h$, normalized to the initial value $h_{\rm initial}$ (blue line), the second derivative of the potential along the solution $|{\rm d}^2 V(h)/{\rm d} h^2|/(3H^2)$ (red line) and the derivative of the field $|{\rm d}\ln h/{\rm d}\ln a|$ (brown line),
as functions of the scale factor $\ln a$.
We have set the parameters as $m_h=125.5$ GeV, $m_t=173.1$ GeV, $H=10^9$ GeV.}
\label{fig:3curves}
\end{figure}

The Higgs field $h$ is initially placed at $h_{\rm initial}$ 
where the second derivative
of the potential is such that $|{\rm d}^2 V(h)/{\rm d} h^2|/(3 H^2)= 10^{-2}$. Then, the field rolls down the potential according to the equation
\be
\ddot h+3 H \dot h+ V'(h)=0\,,
\label{eq:fieldeq}
\ee
where the dot refers to derivative with respect to $t=H^{-1}\ln a$.
We have conventionally set the scale factor equal to 1 at the initial point $h_{\rm intial}$.

In Figure \ref{fig:3curves} we show an example of a situation where the conditions 1. and 2. in the previous section are met and the mechanism works. In fact, the field is slowly rolling down the potential, keeping the second derivative of the potential small (in units of $H^2$), over a wide range
of e-folds. So, in this case it is possible to generate nearly scale-independent isocurvature perturbations of the SM Higgs and, through one of the mechanisms mentioned above,  convert them into 
the observed amount of curvature perturbations 
$P_{\zeta}$.

Next, we repeat the analysis by scanning over
the Higgs and top masses and looking for what values
of $H$ the same situation arises.
Of course, an upper limit on the values of $H$ is set by the instability scale
$\Lambda_{\rm inst}$ at which the Higgs quartic coupling runs negative. For lower values of $H$, down to $\sim 1$ TeV,
we verified that there always exist solutions satisfying
conditions 1  and 2 of the previous section, and generating enough curvature perturbations.
The analysis of the region $H\lesssim 1$ TeV is more delicate  as the Higgs field runs very close to its minimum, but we do not pursue
this here (we are more interested
in the regime where $H$ is relatively large so that tensor modes could be seen by PLANCK).

\begin{figure}[t]
\centering
\includegraphics[scale=1]{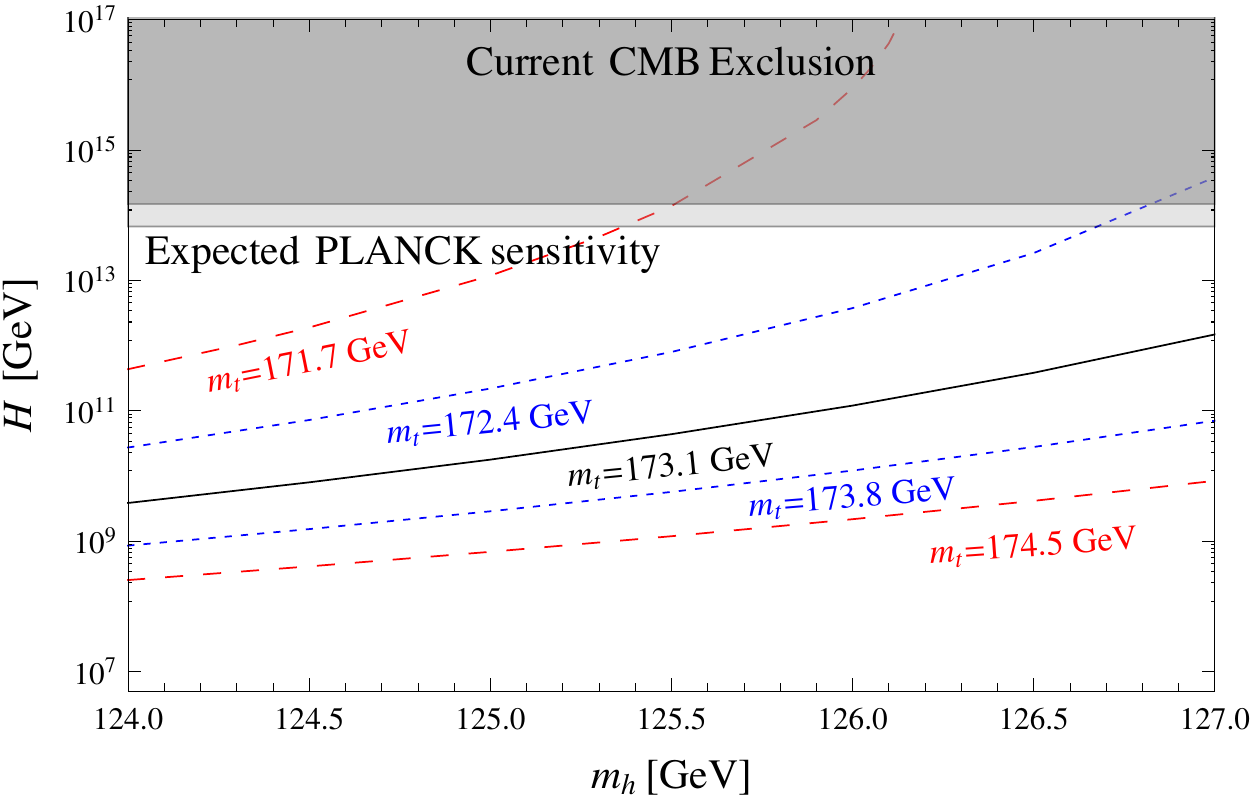}
\caption{Upper limits on $H$ as a function of $m_h$. Solid black line refers to central value of $m_t$, while dotted blue and dashed red lines correspond to $1 \sigma$ and $2\sigma$ variations of $m_t$, respectively.
}
\label{fig:mhvsH}
\end{figure}

In Figure \ref{fig:mhvsH} we show the resulting upper limits on $H$,
corresponding to the instability scale $\Lambda_{\rm inst}$,
for a range of $m_h$ around the observed value.
The different lines refer to $m_t$ at the central value or within 1$\sigma$ and 2$\sigma$ variations.
The mechanism proposed in this paper works for all values of $H$ 
below the curves.
The current exclusion limit on $H$ from CMB data is $H\lesssim 1.5\times 10^{14}$ GeV \cite{wmapping}, which is dark gray band in Fig.~\ref{fig:mhvsH}). As already mentioned, PLANCK data will possibly be able to exclude values of the Hubble rate down to $H\simeq 6.7 \times 10^{13}$ GeV.
This limit is shown as a lighter gray band in Fig.~\ref{fig:mhvsH}. Let us reiterate that the values of $\beta_h$, see Eq. (\ref{aaaa}), is not fixed, but is appropriately
computed at every point to insure that the cosmological perturbation is correctly normalized, $P_\zeta^{1/2}=4.8\times 10^{-5}$. In other words, Figure  2 provides all the possible values of $H$ where the perturbation may be generated by the SM Higgs.

The information in  Figure \ref{fig:mhvsH} can be read in two ways. For  given  $m_h$ and  $m_t$, there is a maximum $H$ for which
the SM Higgs can generate cosmological perturbations, and which
may or may not be in the range to be detected in the near future.
On the other hand, for a given value of $H$, the hypothesis
that the SM Higgs generate perturbations establishes
a correlation between $m_h$ and  $m_t$ which can be tested by
particle physics measurements. 

For example, if the Higgs mass value is confirmed
to be $m_h=125.5$ GeV and $m_t$ and  $\alpha_s$ are at their central values, our mechanism predicts
$H\lesssim 10^{10}$ GeV and therefore tensor perturbations too small
to be detected in the near future. 
On the other hand, if tensor perturbations will be detected
with a given $H$, the mechanism we have proposed makes a prediction for a relation between the Higgs and top masses, 
and we find
\be
(m_h)^{\textrm{B-mode}}\simeq 128.0 \GeV+
1.3\left(\frac{m_t-173.1 \GeV}{0.7 \GeV}\right)  \GeV + 0.9 \left(\frac{H}{10^{15} \GeV}\right) \GeV
\pm \delta_{\rm th}\,,
\label{mhfit}
\ee
where $\delta_{\rm th}\sim 2$ GeV is a residual theoretical uncertainty from higher order corrections,
we have neglected.
The formula (\ref{mhfit}) is valid for $124 \GeV\lesssim m_h\lesssim 127 \GeV$ and
$6.7 \times 10^{13} \GeV\lesssim H \lesssim 1.5\times 10^{14} \GeV$.
The information coming from a more accurate experimental determinations of the SM parameters would then allow to either support or rule out our model.
Notice also that the result in Eq.~(\ref{mhfit}) is crude, and just serves as an illustration. A more careful calculation
(also including 3-loop $\beta$-functions and $\alpha_s$ variation)
would be needed in order to extract a detailed and more complete prediction.

Since the purpose of this paper is to show  whether it is possible to generate cosmological perturbations with the SM Higgs,  the level of accuracy we adopted is enough to conclude that the answer is robustly yes.
A more accurate determination of the Higgs effective potential, using 3-loop $\beta$-functions for gauge, 
Higgs quartic and top-yukawa couplings \cite{3loop} and 2-loop pole mass matching conditions (as recently presented in Refs.~\cite{hi5,Bezrukov:2012sa}),
would be desirable, but is beyond the scope of the present brief communication.

In conclusion, we have pointed out the possibility that the  SM Higgs might be  responsible for the inhomogeneities we observe in our universe, both in the CMB and in the LSS, if there was an early stage of accelerated expansion driven by some vacuum energy and whose size we 
agnostically parametrized with the Hubble rate $H$.  Essentially, the Higgs potential may be flat enough to generate nearly scale-independent isocuravture perturbations which are subsequently converted into the observed adiabatic mode. 
With the recent ATLAS and CMS results for the   Higgs mass, the Higgs can generate the cosmological perturbation in a wide range of the  Hubble rate during inflation,  $H=(10^{10}- 10^{14})$ GeV, 
depending on the values of the SM parameters. On the other side, if the forthcoming data from the Planck satellite will present some hints of a 
$B$-mode in the CMB polarization originated from tensor modes, this will identify a well-defined range of the Higgs mass. 
We have therefore established a very interesting correlation between collider and cosmological measurements,
which makes the  mechanism predictive and falsifiable.

\section*{Acknowledgments}
A.R. is supported by the Swiss National
Science Foundation (SNSF), project ``The non-Gaussian Universe" (project number: 200021140236).

\begin{small}

\end{small}
\end{document}